\documentclass[pra,twocolumn,showpacs]{revtex4}
\usepackage{graphicx}
\begin{document}
\title{Characterization of elastic scattering near a Feshbach resonance in rubidium 87}
\author{Thomas Volz, Stephan D{\"u}rr, Sebastian Ernst, Andreas Marte, and Gerhard Rempe}
\affiliation{Max-Planck-Institut f{\"u}r Quantenoptik, Hans-Kopfermann-Str.\ 1, 85748 Garching, Germany}
\date{\today}
\hyphenation{Fesh-bach}
\hyphenation{pa-ra-bo-la}
\begin{abstract}
The $s$-wave scattering length for elastic collisions between $^{87}$Rb atoms in the state $|f,m_f\rangle=|1,1\rangle$ is measured in the vicinity of a Feshbach resonance near 1007~G. Experimentally, the scattering length is determined from the mean-field driven expansion of a Bose-Einstein condensate in a homogeneous magnetic field. The scattering length is measured as a function of the magnetic field and agrees with the theoretical expectation. The width of the resonance is determined to be 0.20(3)~G, the position of the zero crossing of the scattering length is found at 1007.60(3)~G.
\end{abstract}
%
\maketitle
A Feshbach resonance offers a possibility for tuning the interactions in ultracold atomic gases simply by applying a magnetic field. Feshbach resonances have been used to induce a controlled collapse of a Bose-Einstein condensate (BEC) \cite{donley:01}, to create a coherent superposition of an atomic BEC and a molecular state \cite{donley:02,claussen:condmat0302195}, to realize a bright soliton in a BEC \cite{khayokovich:02,strecker:02}, and to create a BEC of Cs atoms \cite{weber:03}. Various groups are currently trying to use a Feshbach resonance to create a superfluid phase in a degenerate Fermi gas. First results towards this goal have recently been published \cite{ohara:02a}.

Feshbach resonances have been observed in gases of various alkali atoms \cite{inouye:98,courteille:98,roberts:98,vuletic:99,loftus:02,khayokovich:02,strecker:02,jochim:02,dieckmann:02,ohara:02}. With the isotope studied in this paper, $^{87}$Rb, two technical problems must be tackled. One problem is that in $^{87}$Rb Feshbach resonances exist only for internal states that cannot be held in magnetic traps (except for some spin mixtures). Another, more severe problem is that the Feshbach resonances in $^{87}$Rb are unusually narrow: The broadest resonance located near 1007~G is predicted to be only 0.17~G wide. This imposes severe constraints on the current sources used to create the magnetic field. Yet, $^{87}$Rb is the isotope used in the vast majority of today's BEC experiments, thus stimulating a strong interest in investigating the possibility of manipulating the atomic interaction in this isotope. In a recent experiment, we were able to precisely locate more than 40 Feshbach resonances in $^{87}$Rb by observing enhanced atom loss \cite{marte:02}. Similar results for the broadest resonance were obtained very recently at the University of Oxford \cite{foot:pers}.

In the ultracold regime, the elastic scattering properties are fully characterized by the $s$-wave scattering length $a$. Quantitative measurements of $a$ near Feshbach resonances have previously been reported for 3 isotopes:
$^{23}$Na \cite{inouye:98,stenger:99}, $^{85}$Rb \cite{roberts:98,cornish:00,roberts:01a}, and $^{40}$K \cite{loftus:02}. These experiments are based on a measurement of either the mean-field energy of a BEC or the thermalization rate in a non-degenerate gas.

In this paper, the mean-field energy of a BEC is used to measure $a$ in $^{87}$Rb in the vicinity of the Feshbach resonance near 1007~G. Our method differs slightly from the one previously published \cite{inouye:98,stenger:99,cornish:00}. There, the final value of the magnetic field was applied while the atoms were still trapped and the atoms had time to equilibrate before being released from the trap. In this paper, however, we first turn off the trap and then switch to the final value of the magnetic field. This improves the signal-to-noise ratio as discussed at the end of this paper.

The experimental set-up is only briefly summarized here. More details are given in Ref.~\cite{marte:02}. A double-MOT system is used for cooling and trapping of atomic $^{87}$Rb. The atoms are then optically pumped to the state $|f,m_f\rangle=|1,-1\rangle$ and loaded into a Ioffe-Pritchard magnetic trap, in which a radio-frequency (rf) field drives evaporative cooling to the BEC phase transition. It turned out that BECs with the same properties as in Ref.~\cite{marte:02} can be created with a much faster rf sweep than described there. The optimized rf sweep lasts only 5.1~s, resulting in an overall cycle time of 15~s.

After creation of the BEC, the atoms are transferred into an optical dipole trap and held there for roughly 0.5~s. The optical trap is made of two beams from a Nd:YAG laser ($\lambda=1064$~nm), with the beams crossing at right angles. One beam propagates horizontally, along the symmetry axis of the magnetic trap. The second beam subtends an angle of $25^\circ$ with the horizontal plane. The beam waist ($1/e^2$-radius of intensity) and power are $33 ~\mu$m and 38~mW for the horizontal beam; and $77 ~\mu$m  and 115~mW for the second beam. Due to the tighter waist, the horizontal beam creates a stronger confinement than the second beam.

The horizontal beam creates an estimated trap depth in the horizontal plane of $\sim k_B\times 3~\mu$K. In the vertical direction, the trap barely supports the atoms against gravity with a trap depth of $\sim k_B\times 0.8~\mu$K. The horizontal beam creates negligible confinement along its propagation direction. The second beam solves this problem without changing the confinement in the other two directions significantly. The polarizations of the two light beams are orthogonal, so that no optical lattice is formed.

In a coordinate system, where the $z$ axis is the symmetry axis of the magnetic trap and where gravity points along $x$, the trap frequencies are $(\omega_x, \omega_y, \omega_z)= 2\pi \times$ (120, 170, 50)~Hz. The first two frequencies differ due to the gravitational sag. Typically, the atom numbers in the BEC and in the thermal fraction are $10^5$ each. The peak density in the BEC is $\sim 2 \times 10^{14}$~cm$^{-3}$. The optical trap is not designed to maximize the atom number in the BEC. Instead, we try to avoid high peak densities in order to reduce density-dependent losses.

During the transfer from the magnetic trap to the optical trap, a small magnetic bias field ($B \approx 1$~G) preserves the spin polarization of the atoms. After the transfer to the dipole trap, a strong homogeneous magnetic field ($B \approx 1000$~G) is applied in the opposite direction. The rapid switching-on of this field (at rates up to 700~G/ms) flips the atomic spins from state $|1,-1\rangle$ to the absolute ground state $|1,1\rangle$, in which the Feshbach resonance occurs. With this method, the spin-flip efficiency is so close to 100~\%, that a Stern-Gerlach method does not show any significant signal from atoms in wrong spin states. We estimate the detection limit of this measurement to be $\sim 2$~\%.

The 1000-G field is typically not set exactly onto the Feshbach resonance right away. Instead, the field is held a few G above or below the Feshbach resonance for typically 500~ms. During this time, thermal drifts due to the $\sim$12~kW heat load dissipated in the coils have some time to settle. Next, the optical trap is switched off and simultaneously $B$ is jumped to a value very close to or right at the Feshbach resonance. After holding $B$ at its final value for 6~ms, $B$ is switched off completely. The jump to the final field value actually takes $\sim 0.5$~ms, the complete turn-off $\sim 2$~ms. After an expansion time of the order of 20~ms, an absorption image of the expanded cloud is taken with a CCD camera. The atom number and size of the BEC and of the surrounding thermal cloud are determined from a two-dimensional fit to the CCD picture. The magnetic field was calibrated using microwave spectroscopy in the vicinity of the Feshbach resonance with an accuracy of 0.03~G.

\begin{figure}[bt]
\includegraphics{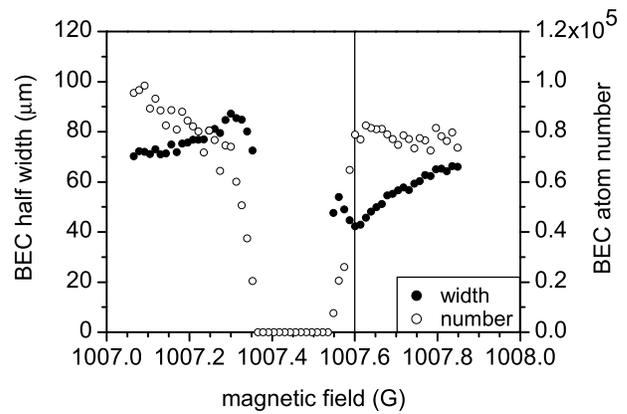}
\caption{
 \label{fig-N-and-W}
Vertical half width ($\bullet$) and atom number ($\circ$) of the expanded BEC as a function of magnetic field $B$. Data points in the left (right) half of the figure were obtained by jumping towards the Feshbach resonance from smaller (larger) $B$, and for an expansion time of 18~ms (23~ms). $B$ was jumped to its final value at the moment of release of the atoms from the trap. $B$ was held there for 6~ms and then switched off. The vertical line at 1007.60~G indicates the onset of instability of the BEC due to negative values of $a$.}
\end{figure}

Figure~\ref{fig-N-and-W} shows the size and atom number of the expanded BEC measured with the timing sequence described above. As expected, the size increases (decreases) as the Feshbach resonance is approached from below (above). Additionally, the atom number is reduced as one approaches the resonance from either side. No BEC is left between 1007.37~G and 1007.53~G. The separation of adjacent data points in Fig.~\ref{fig-N-and-W} is 13~mG which roughly equals the estimated magnetic field resolution due to the measured current noise.

As one gets very close to the Feshbach resonance from either side, the upward (downward) trend in the width is reversed. On the low-field side of the resonance, this is a trivial consequence of the decrease in atom number, because a lower-density BEC releases less mean-field energy that drives the expansion. On the high-field side of the resonance, however, the effect is related to the instability of the BEC in a regime of negative $a$ discussed later in this paper.

In order to extract $a$ from the measured size of the expanded BEC, the expansion process must be modelled. To this end, the model presented in Ref.~\cite{castin:96} is extended to include a possible time dependence of $a$ during the expansion. We start by first ignoring the observed change in atom number. This effect will be incorporated in the analysis later on.

Initially, the BEC is confined in a harmonic potential with trap frequencies $(\omega_x,\omega_y,\omega_z)$. In the Thomas-Fermi approximation to the Gross-Pitaevskii equation, the initial density distribution $n({\bf r})$ is an inverted parabola
\begin{eqnarray}
n({\bf r}) = n_0 \left[ 1- \sum_{k=1}^3 \left(\frac{r_k}{W_k} \right)^2 \right]
\end{eqnarray}
and $n({\bf r}) = 0$ if the above expression is negative. Here, $n_0 = 15 \, N /(8 \pi \, W_x W_y W_z)$ is the peak density. The half widths along the coordinate axes $r_k$ are 
\begin{eqnarray}
\label{eq-W}
W_k(0) = \frac{1}{\omega_k} \left( 15 \; \frac{ \hbar^2 }{m^2} \; \omega_x\omega_y \omega_z \; a_i \, N \right)^{1/5} \; ,
\end{eqnarray}
where $m$ is the atomic mass, $N$ the atom number in the BEC, and $a_i$ the initial scattering length before release. At time $t=0$, the trap is switched off, and a possible time dependence $a(t)$ may begin.

According to Ref.~\cite{castin:96}, the BEC profile stays a parabola during the expansion. Its widths are scaled by the parameters
\begin{eqnarray}
\label{eq-def-lambda}
\lambda_k(t) = \frac {W_k(t)} {W_k(0) } \; .
\end{eqnarray}
The evolution of the scaling parameters $\lambda_k$ is described by a set of coupled differential equations, Eq.~(11) in Ref.~\cite{castin:96}. One can easily show that a time dependence of $a$ during the expansion modifies this equation to
\begin{eqnarray}
\label{diff-eq}
\ddot{ \lambda_k} = \frac{a(t)}{a_i} \; \frac{1}{\lambda_x \lambda_y \lambda_z} \; \frac{\omega_k^2}{\lambda_k} \; .
\end{eqnarray}
The initial conditions are $\lambda_k(0)=1$ and $\dot{\lambda_k}(0)=0$. $a(t)$ appears in Eq.~(\ref{diff-eq}), because the expansion is driven by the mean-field energy which is proportional to $a(t)$. $a_i$ appears in Eq.~(\ref{diff-eq}), because $a_i$ determines the initial widths $W_k(0)$ with respect to which the $\lambda_k$ are defined.

As mentioned earlier, jumping $a$ from its initial to its final value $a_f$ takes $\sim 0.5$~ms. We include this delay in the model in the following way
\begin{eqnarray}
\label{eq-a-t}
a(t) & = & \left\{ 
\begin{array}{ll}
a_{f}  & 0.5 \; {\rm ms} < t < 6 \; {\rm ms}\\ 
a_{i} = a_{\rm bg} \quad & \mbox{otherwise .}
\end{array}
\right.
\end{eqnarray}
Here, $a_{\rm bg}$ is the background value of the scattering length far away from the resonance and $t=0$ is defined by the turn-off of the optical trap. The fact that $a$ is switched back to $a_{\rm bg}$ at $t=6$~ms has little effect, because for the parameters of the experiment almost all mean-field energy is released during the first 6~ms (except for very small values of $a_f$).

The atom loss visible in Fig.~\ref{fig-N-and-W} must also be included in the model. This is fairly easy if it is possible make the approximation that the atom loss is independent of the atomic density. With this approximation, the shape of the BEC remains parabolic during the expansion. With a time dependence of the atom number $N(t)$, Eq.~(\ref{diff-eq}) is then modified to
\begin{eqnarray}
\label{diff-eq-N}
\ddot{ \lambda_k} = \frac{a(t) N(t)}{a_i N_i} \; \frac{1}{\lambda_x \lambda_y \lambda_z} \; \frac{\omega_k^2}{\lambda_k} \; .
\end{eqnarray}
Additionally, in Eq.~(\ref{eq-W}), $N$ is replaced by the initial value $N_i$. Since the dominant loss mechanism is actually likely to be density dependent, this approximation will only be reasonable if a small fraction of the atoms is lost.

It is clear from Eq.~(\ref{diff-eq-N}) that the time dependence of the atom number is crucial. By varying the hold time at the final magnetic field from 1 to 6~ms, we experimentally checked for such a time dependence at $B = 1007.35$~G (last non-vanishing data point on the low-field side in Fig.~\ref{fig-N-and-W}), but found none: The final atom number was independent of the hold time over this range. It is hence obvious that the loss does not occur continuously during the total hold time. Instead, the loss occurs during the first 1~ms of hold time. The origin of this loss might be related to the formation of molecules, but this is not clear. Similar but weaker loss on short timescales was observed for a thermal cloud of atoms in our previous experiment \cite{marte:02}. For simplicity, we assume that the atom loss occurs instantaneously at the same time as the change in $a$
\begin{eqnarray}
\label{eq-N-t}
N(t) & = & \left\{ 
\begin{array}{ll}
N_{f}  & 0.5 \; {\rm ms} < t\\ 
N_{i} \quad & \mbox{otherwise .}
\end{array}
\right. 
\end{eqnarray}

For processing the data, the initial atom number $N_i$ and the background value of the scattering length $a_{\rm bg}$ are needed. $N_i$ was determined from the overall absorption in a measurement far away from the Feshbach resonance. In order to determine $a_{\rm bg}$, the size of an expanded BEC far away from the Feshbach resonance was measured. Using Eqs.~(\ref{eq-W})-(\ref{diff-eq}) with $a(t)=a_i=a_{\rm bg}$, the background scattering length can be extracted, yielding $a_{\rm bg} = 108 (30) \, a_0$, where $a_0$ is the Bohr radius. This is consistent with the theoretical value $a_{\rm bg} = 100.5 \, a_0$ for the state $|1,1\rangle$ \cite{varhaar:pers}. For further data processing, the theoretical value was used.

Given $N_i$ and $a_{\rm bg}$, $W_x(0)$ is calculated using Eq.~(\ref{eq-W}), and thus the measured expanded widths $W_x(t)$ can be converted into $\lambda_x(t)$ according to Eq.~(\ref{eq-def-lambda}). For every data point, the observed final atom number $N_f$ in the BEC was then used to numerically solve the coupled differential equations (\ref{diff-eq-N}) for the experimentally applied sequence Eqs.~(\ref{eq-a-t}), (\ref{eq-N-t}). In the calculation, $a_f$ was varied until the experimentally observed width was matched.

\begin{figure}[bt]
\includegraphics{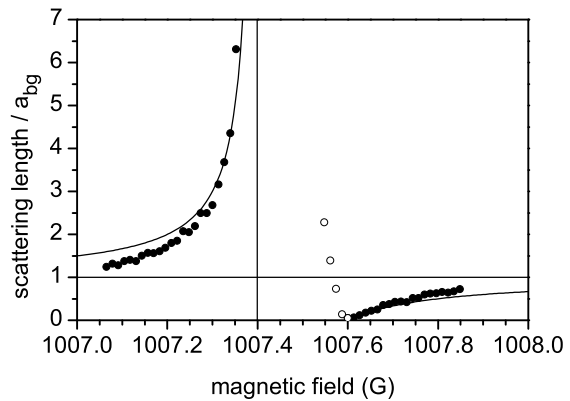}
\caption{
 \label{fig-a-vs-B}
Scattering length as a function of magnetic field. The solid line is a fit to the experimental data ($\bullet$). The best fit yields $\Delta B=0.20$~G for the width and $B_{\rm peak}=1007.40$~G for the position of the resonance. Some experimental data points ($\circ$) were not included in the fit, because they lie in the regime of negative $a$ where the BEC is unstable.}
\end{figure}

With this method, the data shown in Fig.\ref{fig-N-and-W} were processed to extract the scattering length, which is shown in Fig.~\ref{fig-a-vs-B}. Figure~\ref{fig-a-vs-B} also displays a fit to the theoretical expectation (see e.\ g.\ \cite{timmermans:99})
\begin{eqnarray}
a = a_{\rm bg} \left(1 - \frac{\Delta B} {B - B_{\rm peak}} \right) \; .
\end{eqnarray}
Here, $\Delta B$ is the width of the Feshbach resonance and $B_{\rm peak}$ the position of the pole in $a$.

Some data points ($\circ$ in Fig.~\ref{fig-a-vs-B}) are not included in the fit for the following reason: Approaching the resonance from above in Fig.~\ref{fig-N-and-W}, one can see that the size of the expanded BEC decreases as expected, until at 1007.60~G (vertical line in Fig.~\ref{fig-N-and-W}) this trend is reversed. At the same field, significant atom loss suddenly begins. We interpret this as the onset of instability of the BEC in a regime of negative scattering length \cite{bradley:97,sackett:99,gerton:00,roberts:01,donley:01,chin:condmat0212568}. This interpretation is further supported by the fact that the extracted value of $a$ shown in Fig.~\ref{fig-a-vs-B} reaches zero at this field. For the parameters of our experiment, the critical scattering length for instability \cite{roberts:01,claussen:condmat0302195} is $a_{\rm crit} \approx -10^{-3} \, a_{\rm bg}$ which is almost zero. Hence, we conclude that the zero crossing of $a$ is located at $B_{\rm zero} = B_{\rm peak} + \Delta B=1007.60 (3)$~G and we use only one free parameter $\Delta B$ in our fit. A more detailed analysis of the region of instability is clearly beyond the scope of this paper.

The fit in Fig.~\ref{fig-a-vs-B} is in reasonable agreement with the experimental data. Data points far away from the resonance are systematically too close to unity. This might be due to the fact that the magnetic field was not held infinitely far away from the resonance before jumping to the resonance. The best-fit value for the width is $\Delta B = 0.20(3)$~G, resulting in $B_{\rm peak} = 1007.40(4)$~G. These values are consistent with the theoretical predictions $\Delta B^{\rm theory} = 0.17(3)$~G and $B_{\rm peak}^{\rm theory} = 1008.5(1.6)$~G \cite{marte:02}. $B_{\rm peak}$ is also near the peak of the atom loss at $1007.34(3)$~G measured in Ref.~\cite{marte:02}.

As mentioned in the introduction, the method to determine $a$ described in this paper differs slightly from the method published previously \cite{inouye:98,stenger:99,cornish:00}. There, the final magnetic field was applied while the BEC was still in the trap and the system had time to equilibrate before the BEC was released from the trap. No matter if the expansion time is almost zero \cite{cornish:00} or long \cite{inouye:98,stenger:99} (in which case $B$ was still on during the initial expansion), the observed BEC size $W$ yields $a \propto W^5$. With the method used in this paper, however, $a \propto W^2$. This is because here the initial BEC peak density $n_i$ is independent of $B$. Hence, the mean-field energy ($\propto n_i a$) is converted into kinetic energy, so that the final velocity (and thus $W$) is proportional to $a^{1/2}$. Therefore, with our method noise in the determination of $W$ is not amplified as much when extracting $a$. A similar technique was used in Ref.~\cite{weber:03}, but there no quantitative values of $a$ were extracted.

To summarize, the mean-field driven expansion of a BEC was used to measure the scattering length $a$ in $^{87}$Rb as a function of the magnetic field in the vicinity of the Feshbach resonance near 1007~G. The position and width of the resonance were extracted from the data. Despite the fact that the resonance is very narrow, we have clearly demonstrated the variation of $a$ over a wide range. The range explored here is currently not limited by magnetic field noise, but rather by fast atom loss near the resonance.

This work was supported by the European-Union Research-Training Network ``Cold Quantum Gases" and the German-Israeli Foundation for Scientific Research and Development.


\end{document}